\documentclass[twocolumn,showpacs,pra]{revtex4}
\usepackage{amssymb}
\usepackage{graphicx}
\usepackage{dcolumn}
\usepackage{bm}
\usepackage{amsmath}
\usepackage{epsfig}

\begin{document}

\title{Deflection of Rotating Symmetric Molecules by Inhomogeneous Fields}

\author{E. Gershnabel}
\author{I.Sh. Averbukh}
\affiliation{Department of Chemical Physics, The Weizmann Institute
of Science, Rehovot 76100, ISRAEL}

\begin{abstract}

We consider deflection of rotating symmetric molecules by inhomogeneous optical and static electric fields, compare results with the case of linear molecules, and find new singularities in the distribution of the scattering angle. Scattering of the prolate/oblate molecules is analyzed in detail, and it is shown that the process can be efficiently controlled  by means of short and strong femtosecond laser pulses. In particular, the angular dispersion of the deflected molecules may be dramatically reduced by laser-induced molecular pre-alignment. We first study the problem by using a simple classical model, and then find similar results by means of more sophisticated methods, including the formalism of adiabatic invariants  and direct numerical simulation of the Euler-Lagrange equations of motion. The suggested control scheme opens new ways for many applications involving molecular focusing, guiding, and trapping by optical and static fields.

\end{abstract}


\maketitle

\section{Introduction} \label{Introduction}

Formation of aligned or oriented molecules has long been of interest in chemistry and physics. Modern applications of aligned and oriented molecules, such as high-harmonic generation \cite{Wayne}, laser pulse compression \cite{Bartels}, nanolithography \cite{Gordon}, control of photodissociation and photoionization \cite{Larsen}, and quantum information processing \cite{Shapiro}, have motivated the development of all-optical techniques for aligning molecules under field-free conditions. A major advance has been the use of linearly polarized, nonresonant ultrashort laser pulses to align molecules by an impulsive Raman mechanism \cite{Heritage}. Short laser pulses excite rotational wave packets, which results in a considerable transient molecular alignment after the laser pulse is over, i.e. at field-free conditions (for recent studies of laser induced field-free alignment of non-linear molecules, see e.g. \cite{Stapelfeldt_symm,Hamilton_symm,Kitano,Pabst}). These methods also provided tools for more general control of the molecular dynamics. In particular, we have recently shown that laser induced field-free alignment may affect dramatically the scattering of linear molecules from inhomogeneous optical and static electrical fields \cite{Gershnabel,Gershnabel1,Floss}. This opens new ways for many applications in molecular optics, involving molecular focusing, guiding and trapping by optical and static fields \cite{Stapelfeldt,Bum,Purcell}.

As most  molecules are not linear,  a more general theory is required for describing scattering of symmetric and asymmetric molecules by inhomogeneous fields. In this work, we focus on the  scattering of symmetric molecules by optical and static electrical fields, and investigate distinctive features in the distribution of scattering angles due to molecular rotation. In particular, we investigate rotational rainbows in scattering, and find a new one in addition to those described in our previous study on the scattering of linear molecules  \cite{Gershnabel}. The difference between the symmetric and the linear molecules can be easily seen at the classical level. Having three non-vanishing moments of inertia, symmetric molecules rotate in a more complicated fashion \cite{Landau}, and unlike the linear molecules, their rotation is not constrained to a plane perpendicular to the angular momentum. We will show that the scattering of symmetric molecules can be significantly affected and controlled by preshaping molecular angular distribution before the molecules enter the interaction zone. This can be done with the help of numerous recent techniques for laser molecular alignment, which use single or multiple short laser pulses (transform limited or shaped) to align molecular axes along certain directions.
This paper is based on classical considerations, since we have demonstrated in the past a good correspondence between the quantum mechanical and classical treatments  \cite{Gershnabel,Gershnabel1} of the strong-field rotational control. The first part of the paper, i.e. Secs. \ref{Deflectoni of Molecules}-\ref{Control of the scattering of symmetric molecules} are devoted to the molecular deflection by optical fields. The last part of the paper, i.e. Sec. \ref{Static Deflection of Symmetric Molecules} describes the deflection of symmetric molecules by a static electric field. In Sec. \ref{Deflectoni of Molecules} we introduce the general description of the molecular deflection  by optical fields. In Sec. \ref{Simple Model}, a simple model of a symmetric molecule is presented, for which rotational rainbows are predicted in the distribution of the scattering angle. In Sec. \ref{Thermal ensemble} we generalize the simplified model to thermal conditions. In Sec. \ref{real molecules: oblate vs. prolate}, we consider examples of real molecules, and compare between the oblate and prolate cases. In Sec. \ref{Control of the scattering of symmetric molecules}, laser-induced field-free alignment (for various polarizations) is implemented in order to control the scattering process. We expand the study to deflection by a static electric field in Sec. \ref{Static Deflection of Symmetric Molecules}, and finally, we summarize in Sec. \ref{Conclusions}.

\section{Deflection of Molecules} \label{Deflectoni of Molecules}

Although our arguments are rather general, we follow for certainty a
deflection scheme similar to one of the experiment by Stapelfeldt $et$
$al$ \cite{Stapelfeldt} who used a strong IR laser to deflect beams of
linear molecules, and then addressed a portion of the
deflected molecules (at a preselected place and time) by an
additional short and narrow ionizing pulse. Consider  deflection (in
$Z$ direction) of a  molecule moving in $X$ direction  with
velocity $v_X$ and interacting with a focused nonresonant laser beam
that propagates along the $Y$ axis.

The spatial profile of the laser electric field in the $XZ$-plane
is:
\begin{equation}
E=E_0\exp [-(X^2+Z^2)/\omega_0^2 ]\exp [-2\ln2t^2/\tau^2
].\label{E_deflect_field}
\end{equation}

We consider a laser field polarized in the $Z$ direction, and find the interaction potential of the laser with the asymmetric molecules to be \cite{Seideman}:
\begin{equation}
U_a=-\frac{1}{4}E^2\left(\alpha^{ab}\cos^2\theta+\alpha^{cb}\sin^2\theta\sin^2\chi+\alpha_b\right),\label{Interaction_asymmetric}
\end{equation}
where $E$ is defined in Eq. \ref{E_deflect_field}, and $\theta$ and $\chi$ are the Euler angles. $\alpha_{a,b,c}$ are the polarizability components of the molecule, along its principal axes ($a,b,c$), and $\alpha^{ij}\equiv \alpha_i-\alpha_j$. Here, the Euler angles, as well as the principal axes of the molecule are defined as in Fig. $2$ of Ref. \cite{water Gershnabel}. In the case of symmetric molecules, we can define the parallel and perpendicular polarizability components ($\alpha_\parallel$ and $\alpha_\perp$, respectively) as: $\alpha_a\equiv \alpha_{\parallel}$ and $\alpha_b=\alpha_c\equiv \alpha_{\perp}$. Using Eq. \ref{Interaction_asymmetric}, one obtains the interaction term:

\begin{equation}
U_s=-\frac{1}{4}E^2\left [(\alpha_\parallel-\alpha_\perp)\cos^2\theta+\alpha_\perp\right],\label{Interaction_symmetric}
\end{equation}
which has the same form as for the linear molecules.

A molecule initially moving along the $X$ direction
 acquires a velocity component $v_Z$ along $Z$-direction. We
consider the perturbation regime corresponding to a small deflection
angle, $\gamma\thickapprox v_Z/v_X$. We  treat $Z$ as a fixed impact parameter,
and substitute $X=v_X t$. By doing this, we concentrate on the molecules reaching the
focal spot at the moment of the maximum of the deflecting pulse, like in Refs. \cite{Stapelfeldt,Purcell}.  The deflection velocity is given by:

\begin{equation}\label{Velocity_Deflection}
v_Z = \frac{1}{M}\int_{-\infty}^{\infty}F_Z dt
=-\frac{1}{M}\int_{-\infty}^{\infty}\left(\overrightarrow{\nabla}U_s\right)_Z
dt.
\end{equation}

Here $M$ is the mass of the molecules, and $F_Z$ is the deflecting
force. The time-dependence of the force $F_Z$ (and potential $U_s$) in
Eq.(\ref{Velocity_Deflection}) comes from three sources: pulse
envelope, projectile motion of the molecule through the laser focal
area, and time variation of the angle $\theta$ due to molecular
rotation. For simplicity, we start with the case of a relatively
weak deflecting field that does not affect significantly the
rotational motion. Such approximation is justified, say for Benzene
molecules with the rotational temperature $T=5K$, which are subject
to the deflecting field of $3\cdot10^9 W/cm^2$. The corresponding
alignment potential
$U\approx-\frac{1}{4}\left(\alpha_\parallel-\alpha_\perp\right)E_0^2\approx
0.02\ meV$  is an order of magnitude smaller than the thermal energy
$k_BT$, where $k_B$ is Boltzmann's constant. This assumption is even
more valid if the molecules were additionally subject to the
aligning pulses prior to deflection.

Since the rotational time scale is the shortest one in the problem, we average  the force over the fast rotation, and arrive at the following expression for the deflection angle, $\gamma = v_Z/v_X$:
\begin{equation}
\gamma = \gamma_0 \left[(\alpha_\parallel-\alpha_\perp){\cal A}+\alpha_\perp\right].\label{Deflection Angle symmetric}
\end{equation}
Here ${\cal A}\equiv\overline{\cos^2\theta}$ denotes the time-averaged value of
$\cos^2\theta$. This quantity is different for different molecules of the
incident ensemble, which leads to the randomization of the
deflection process. The constant $\gamma_0$ presents the average
deflection angle for an isotropic molecular ensemble:
\begin{eqnarray}
\gamma_0 &=&\frac{E_0^2}{4Mv_X^2}\left(\frac{-4Z}{\omega_0}\right) \nonumber\\
&\times&\sqrt{\frac{\pi}{2}}\left(1+\frac{2\omega_0^2\ln2}{\tau^2v_{X}^2}\right)^{-1/2}\exp\left(-\frac{2Z^2}{\omega_0^2}\right).\label{Average_Deflection Angle asymmetric}
\end{eqnarray}

\section{Simple Model} \label{Simple Model}

\begin{figure}[htb]
\begin{center}
\includegraphics[width=4cm]{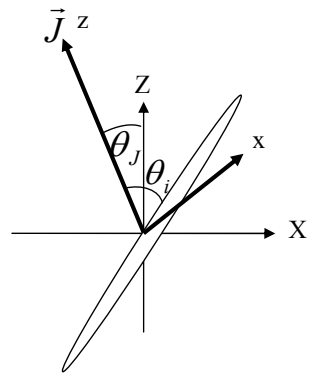}
\end{center}
\caption{The symmetric molecule model. $X,Y,Z$ are the laboratory coordinates, and $\theta_J$ is the angle between the angular momentum $\vec{J}$ and the $Z$ axis. The $x,y,z$ coordinates are obtained by rotating the $X,Y,Z$ coordinates about $Y$ by an angle $\theta_J$, such that the $z||\vec{J}$. The molecule precesses about its own axis, and the axis itself rotates in a conical trajectory about $\vec{J}$.} \label{SymmetricModel}
\end{figure}

We consider a symmetric rotor, as in Fig. \ref{SymmetricModel} (the coordinates are defined in the figure). This rotor precesses about its own axis, and the axis itself rotates about the angular momentum $\vec{J}$ (the angle $\theta_i$ between the molecular axis and $\vec{J}$ remains constant).
As  seen from Eq. \ref{Interaction_symmetric}, the time-averaged interaction of the symmetric molecule and the laser field is linearly related with ${\cal A}$ (${\cal
A}\equiv\overline{\cos^2\theta}$, as was defined before). In what follows, we will estimate the magnitude of ${\cal A}$. In the $x,y,z$ coordinates, the molecular axis direction is given by:
\begin{equation}
\hat{r}_{x,y,z}=(\sin\theta_i\cos\omega t,\sin\theta_i \sin\omega t,\cos\theta_i),
\label{molecular axis,x,y,z}
\end{equation}
where $\omega$ is the frequency of rotation. The relation between the $x,y,z$ and the $X,Y,Z$ coordinates is:
\begin{equation}
\left( \begin{array}{c}
X  \\
Y  \\
Z  \end{array} \right)
=\left( \begin{array}{ccc}
\cos\theta_J & 0 & -\sin\theta_J \\
0 & 1 & 0 \\
\sin\theta_J & 0 & \cos\theta_J \end{array} \right)
\left( \begin{array}{c}
x  \\
y  \\
z  \end{array} \right).
\label{xyz to XYZ}
\end{equation}
Therefore, the molecular axis in the $X,Y,Z$ coordinates  is:
\begin{equation}
\hat{r}_{X,Y,Z}=\left( \begin{array}{ccc}
\cos\theta_J & 0 & -\sin\theta_J \\
0 & 1 & 0 \\
\sin\theta_J & 0 & \cos\theta_J \end{array} \right)
\hat{r}_{x,y,z}\label{molecular axis,X,Y,Z}
\end{equation}
and the projection of the molecular axis on the $Z$ axis is given by:
\begin{equation}
\hat{r}_{X,Y,Z}\cdot\hat{Z}=\cos\theta=\sin\theta_J\sin\theta_i\cos\omega t+\cos\theta_J\cos\theta_i,
\label{projection, symmetric}
\end{equation}
where $\theta$ is the angle between the molecular axis and the $Z$ axis. Taking the square, and averaging it over time, we obtain:
\begin{equation}
{\cal A}=\frac{1}{2}\sin^2\theta_J\sin^2\theta_i+\cos^2\theta_J\cos^2\theta_i.
\label{A symmetric}
\end{equation}
To find singularities in the distribution of scattering angles, we determine the extremal points $\frac{\partial {\cal A}}{\partial \theta_{i,J}}=0$, and find that:
\begin{equation}
\frac{\partial {\cal A}}{\partial \theta_{i,J}}=0 \rightarrow \left\{ \begin{array}{lll}
         \sin2\theta_{i}=0 &and &\sin2\theta_{J}=0\\
         &or&\\
        \sin^2\theta_{i}=\frac{2}{3}&and & \sin^2\theta_{J}=\frac{2}{3}.\end{array} \right.
\label{derivatives symmetric}
\end{equation}

According to the first line of Eq. \ref{derivatives symmetric}, the extremal angles $\theta_i$ and $\theta_J$  can be either $0,\pi/2,\pi$. Considering smooth distributions of the angles $\theta_i ,\theta_J $, the regions near $0,\pi$ give a negligible contribution to the distribution of ${\cal A}$ because of the vanishing phase space volume.
Therefore we conclude that the distribution of ${\cal A}$  may have singularities at:
\begin{eqnarray}
&&{\cal A}(\theta_i=\pi/2,\theta_J=\pi/2)=1/2\nonumber\\
&&{\cal A}(\sin^2\theta_i=2/3,\sin^2\theta_J=2/3)=1/3.
\label{singularities symmetric}
\end{eqnarray}

\section{Scattering of a Thermal ensemble} \label{Thermal ensemble}

Consider a symmetric rigid rotor described by a Lagrangian:
\begin{equation}
L=\frac{P_\theta^2}{2I_c}+\frac{\left(  P_\phi-P_\chi\cos\theta  \right)^2}{2I_c\sin^2\theta}+\frac{P_\chi^2}{2I_a},
\label{symmetric lagrangian}
\end{equation}
where $I_b=I_c$, and $I_a$ is the moment of inertia about the molecular symmetry axis. $P_\theta$, $P_\phi$ and $P_\chi$ are the canonical momenta. The angular momentum $J$ can be found by:
\begin{equation}
J^2=P_\chi^2+P_\theta^2+\frac{\left( P_\phi-P_\chi\cos\theta  \right)^2}{\sin^2\theta}.
\label{angular momentum}
\end{equation}
Using this equation, and also the expression for the total energy defined by Eq. \ref{symmetric lagrangian}, one can
easily find the angles $\theta_i , \theta_J$ introduced before:
\begin{eqnarray}
\sin\theta_i&=&\sqrt{\frac{\frac{2I_c E}{J^2}-\frac{I_c}{I_a}}{1-\frac{I_c}{I_a}}}\nonumber\\
\cos\theta_J&=&\frac{P_\phi}{J}.
\label{calculate model angles}
\end{eqnarray}

For a thermal molecular ensemble with temperature $T$, we define dimensionless momenta $P_\phi '=\frac{P_\phi-P_\chi\cos\theta}{\sqrt{I_c K_B T}\sin\theta}$, $P_\chi '=\frac{P_\chi}{\sqrt{I_a K_B T}}$ and $P_\theta '=\frac{P_\theta}{\sqrt{I_c K_B T}}$, and  calculate the distribution of the time-averaged alignment factor ${\cal A}$ as:
\begin{eqnarray}
f({\cal A})&=&\int\int\int\int\int d\theta(0)\sin\theta(0)  d\phi(0) d\chi(0)   \nonumber\\
&\times&dP_\theta '(0) dP_\chi ' (0) dP_\phi '(0)\delta({\cal A}-\overline{\cos^2\theta})\nonumber\\
&\times& F(\theta(0),\phi(0),P_\theta '(0),P_\chi '(0),P_\phi ' (0)),
\label{probability distribution equation}
\end{eqnarray}
where
\begin{equation}
F=\frac{1}{Q_{rot}}\exp \left [ -\frac{1}{2}\left( P_\chi'^2+P_\theta'^2 +P_\phi'^2  \right)\right],
\label{F of symmetric}
\end{equation}
and $Q_{rot}$ is the partition function. Here $\overline{\cos^2\theta}$ is given by Eq. \ref{A symmetric}.

\begin{figure}[htb]
\begin{center}
\includegraphics[width=4cm]{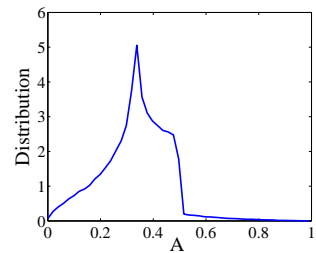}
\end{center}
\caption{Distribution of ${\cal A}$ for a symmetric molecule. $T=5K$, and the following values of the moments of inertia were used: $I_a=1\cdot 10^{-45}kg\cdot m^2$,$I_b=2\cdot 10^{-45}kg\cdot m^2$ and $I_c=2\cdot 10^{-45}kg\cdot m^2$. The calculations are based on Eq. \ref{A symmetric}, and a Monte-Carlo simulation.} \label{Results A symmetric}
\end{figure}

A Monte-Carlo simulation was performed, in order to account for the random initial orientation and momenta of the molecules in the thermal ensemble. The results are shown in Fig. \ref{Results A symmetric}, where we see singularities at the points ${\cal A}=1/3,1/2$, in full agreement with the previous speculations.

We have also verified the results presented in Fig. \ref{Results A symmetric} by using two other methods. First, based on the separation of the rotational and translational time scales, we used adiabatic invariant technique \cite{Landau,Goldstein,Dugourd} to calculate ${\cal A}$ for a symmetric molecule moving in inhomogeneous field, and utilized these data for calculating the deflection angles.  In the second method, we numerically solved the Euler-Lagrange equation of motion, calculated $\cos^2\theta(t)$, and averaged it over a long enough time to reach convergence \cite{Gershnabel1}. In both methods, the final results were similar to those shown in Fig. \ref{Results A symmetric}.

\section{real molecules: oblate vs. prolate} \label{real molecules: oblate vs. prolate}

\begin{figure}[htb]
\begin{center}
\includegraphics[width=4cm]{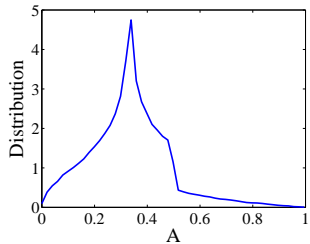}
\end{center}
\caption{Distribution of ${\cal A}$, for the Benzene molecule. The calculations are based on Eq. \ref{A symmetric}.} \label{BenzeneA}
\end{figure}

We are ready now to apply the developed procedure to real molecules. We  start with Benzene, i.e. an oblate molecule. For this molecule, the symmetry axis is $a$, and $I_a=2\cdot I_b$, where $I_b=I_c=1.474\cdot 10^{-45} kg\cdot m^2$ (moments of inertia and polarizability data are taken from \cite{Hasegawa}). The distribution of ${\cal A}$ for a Benzene molecule is given in Fig. \ref{BenzeneA}. It is seen that the  distribution peak near ${\cal A}=1/3$  is the dominant one in the case of Benzene.

\begin{figure}[htb]
\begin{center}
\includegraphics[width=4cm]{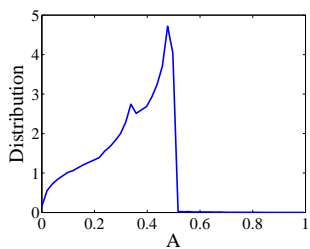}
\end{center}
\caption{Distribution of ${\cal A}$, for the Methyl Iodide ($CH_3I$) molecule. The calculations are based on Eq. \ref{A symmetric}.} \label{MethylA}
\end{figure}

As another example, we consider Methyl Iodide ($CH_3I$), which is a prolate symmetric molecule. Again, the symmetry axis is $a$, but the moments of inertia are: $I_a=5.4071\cdot 10^{-47} kg\cdot m^2$, $I_b=I_c=1.118\cdot 10^{-45} kg\cdot m^2$ (moments of inertia data are taken from \cite{Hartinger}, and polarizability data are taken from \cite{Ito}). The corresponding distribution is presented in Fig. \ref{MethylA}. For this molecule, the most visible peak in the distribution is around the ${\cal A}=1/2$ point (quite similar to the case of  linear molecules \cite{Gershnabel}).

\section{Controlling  the scattering of symmetric molecules by a laser field} \label{Control of the scattering of symmetric molecules}

\begin{figure}[htb]
\begin{center}
\includegraphics[width=2cm]{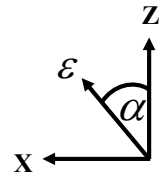}
\end{center}
\caption{The deflecting field is in the laboratory $Z$ direction, and the prealignment pulse (denoted as $\varepsilon$)  is tilted by an angle $\alpha$ with respect to it.} \label{prealignmentDiagram}
\end{figure}

Here we consider the laser deflection of symmetric molecules, that are prealigned by a femtosecond laser pulse before entering the deflecting field. The polarization of the prealignment laser pulse is not necessarily in the direction of the deflecting field, as is shown in Fig. \ref{prealignmentDiagram}.
After the end of the prealigning pulse, the quantities $P_\theta(0)$ and $P_\phi(0)$ are replaced by:
\begin{eqnarray}
P_\theta(0)&\rightarrow& P_\theta(0)+P_s (  \sin^2\alpha\sin 2\theta\cos^2\phi\nonumber\\
&&+\sin 2\alpha\cos 2\theta \cos\phi-\sin 2\theta\cos^2\alpha)\nonumber\\
P_\phi(0)&\rightarrow& P_\phi(0)-P_s(\sin^2\alpha\sin^2\theta\sin 2\phi\nonumber\\
&&+\frac{1}{2}\sin 2\alpha \sin 2 \theta \sin\phi),
\label{changeMomentaSymmetric}
\end{eqnarray}
where $P_s=P\hbar$ and $P=(1/4\hbar)(\alpha_\parallel-\alpha_\perp)\int\varepsilon^2(t)dt$. The dimensionless parameter $P$ is the so called "kick strength" that equals to the typical amount of angular momentum, in the units of $\hbar$, supplied by the pulse to the molecule. For Benzene, $\Delta\alpha = \alpha_\parallel-\alpha_\perp$ is a negative quantity, therefore $P$ is negative.

\begin{figure}[htb]
\begin{center}
\includegraphics[width=8cm]{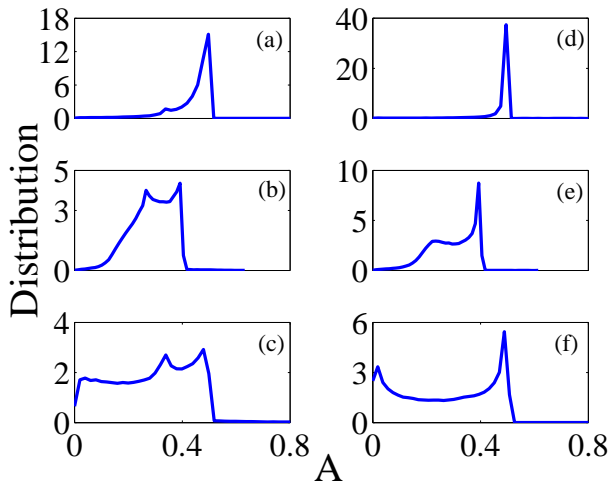}
\end{center}
\caption{Benzene ((a)-(c))and $CH_3I$ ((d)-(f)) ${\cal A}$ distributions, after the ensembles were prealigned by the means of a fs laser pulse, at different angles $\alpha$ ((a) and (d) correspond to $\alpha=0^\circ$; (b) and (e) correspond to $\alpha=45^\circ$; (c) and (f) correspond to $\alpha=90^\circ$). $T=5K$, $P=-25,25$ for Benzene and $CH_3I$, respectively. The right column (Prolate) is very similar to the linear molecule, where the left column (Oblate) shows the peak at $1/3$, in addition.} \label{Symmetric with pulse}
\end{figure}

We analyzed  the effect of prealigning pulses on the scattering of two molecules considered in the previous section. The kick strength was taken as large as $|P|=25$. This corresponds to a prealignment pulse with a peak intensity of $4.13\cdot 10^{12}{W/cm^2}$ and a duration of $0.5ps$ (FWHM) for Benzene. For $CH_3I$,  $P=25$ corresponds to the pulse  intensity of $2.41\cdot 10^{14} W/cm^2$ and a duration of $20fs$ (FWHM). Fig. \ref{Symmetric with pulse} shows distribution of ${\cal A}$  both for Benzene and $CH_3I$ , after the molecules were prealigned by a laser pulse tilted at different angles $\alpha$ ($0^\circ, 45^\circ$ and $90^\circ$) with respect to the deflecting field.
In the prolate molecule case, the distribution is somehow similar to the distribution of a linear rotor that was kicked by the prealignment pulse. When the prealignment polarization is parallel to the deflecting field ($\alpha=0^\circ$), the vector of the angular momentum of the molecules is preferentially confined to the $XY$ plane after the pulse, i.e. $\theta_i\approx\theta_J\approx\pi/2$, which corresponds to ${\cal A}=1/2$. In this way, the molecules experience the maximally possible time-averaged deflecting force which is the same for all the particles of the ensemble. As the result, the dispersion of the scattering angles is reduced dramatically (Fig. \ref{Symmetric with pulse}, $CH_3I,\alpha=0$) \cite{Gershnabel}. For other polarization angles, two peaks are observed i.e. at  ${\cal A}=1/2$ and $(\cos^2\alpha)/2$ (Fig. \ref{Symmetric with pulse}, $CH_3I,\alpha=45,90$) \cite{Floss}. For oblate molecules, the difference from  the linear molecules is more emphasized due to the dominant peak at ${\cal A}=1/3$.

\section{Deflection of Symmetric Molecules by Static Electric Fields} \label{Static Deflection of Symmetric Molecules}
\begin{figure}[htb]
\begin{center}
\includegraphics[width=5cm]{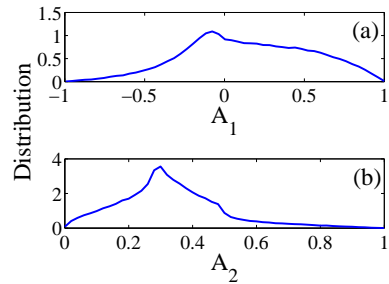}
\end{center}
\caption{Statistical distributions of (a) ${\cal A}_1$ and (b) ${\cal A}_2$ for a thermal beam of $CF_3H$ molecules in a moderate deflecting field ($2\times 10^7 V/m$). $T=10K$, $C=-9.7\times 10^{-5}$ and $D=0.8$.} \label{StaticSmallField_NoPre}
\end{figure}

\begin{figure}[htb]
\begin{center}
\includegraphics[width=5cm]{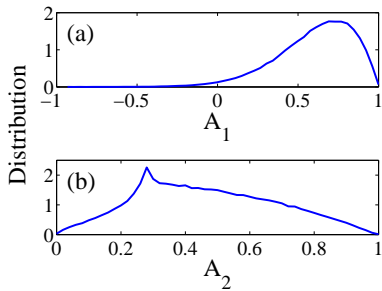}
\end{center}
\caption{Statistical distributions of (a) ${\cal A}_1$ and (b) ${\cal A}_2$ for a thermal beam of $CF_3H$ molecules in a strong deflecting field ($2\times 10^8 V/m$). $T=10K$, $C=-9.7\times 10^{-3}$ and $D=8$.} \label{StaticHighField_NoPre}
\end{figure}

\begin{figure}[htb]
\begin{center}
\includegraphics[width=5cm]{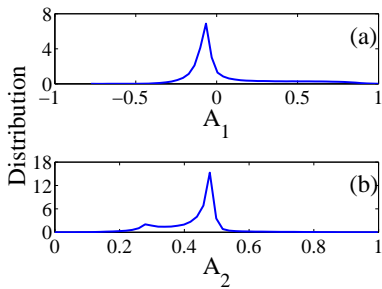}
\end{center}
\caption{Statistical distributions of (a) ${\cal A}_1$ and (b) ${\cal A}_2$ for a prealigned ($P=25$) thermal beam of $CF_3H$ molecules in a strong deflecting field ($2\times 10^8 V/m$). $T=10K$, $C=-9.7\times 10^{-3}$ and $D=8$.} \label{StaticHighField_Pre25}
\end{figure}

In this section we expand the research to scattering of symmetric molecules by static inhomogeneous electric fields. We follow here the same line of thought presented in our recently published paper \cite{Gershnabel1}, where we considered a deflection by a static electric field of a beam of linear molecules. Briefly, in the case of deflection by a static electric field, the field interacts not only with the molecular polarizability, but also with its permanent dipole moment. Thus, Eq. \ref{Interaction_symmetric} is modified as:
\begin{equation}
U_s=-\frac{1}{2}E^2\left [(\alpha_\parallel-\alpha_\perp)\cos^2\theta+\alpha_\perp\right]-\mu E \cos\theta.
\label{Interaction_symmetric_static}
\end{equation}
According to Sec. \ref{Deflectoni of Molecules}, the important variables  defining the deflection are given by the time-averaged values of $\overline{\cos\theta}$ and $\overline{\cos^2\theta}$, i.e.
\begin{eqnarray}
{\cal A}_1&=&\overline{\cos\theta}\nonumber\\
{\cal A}_2&=&\overline{\cos^2\theta}.\label{Simple_COS_avg_time}
\end{eqnarray}
As in \cite{Gershnabel1}, we estimate the strength of the deflecting field by means of the following parameters: $C\equiv E^2\Delta\alpha/(k_B T)$ and $D\equiv \mu E/(k_B T)$. For instance, for a $10K$ $CF_3H$ molecular beam that is deflected by a static field of $2\times 10^7 V/m$,  the corresponding parameter values are $C=-9.7\times 10^{-5}$ and $D=0.8$ (the typical electrical properties of the $CF_3H$ molecule were taken from \cite{Meerts}). In this case, the interaction with the induced polarization may be neglected. We consider only the interaction with the permanent dipole moment of the molecule, and calculate ${\cal A}_1$ and ${\cal A}_2$ by means of the adiabatic invariants method (the derivation is given in \cite{Dugourd}). The results are presented in Fig. \ref{StaticSmallField_NoPre}. The distribution of the deflection angle is linearly related with the distribution of ${\cal A}_1$ (Fig. \ref{StaticSmallField_NoPre}a), however we provide also the distribution of ${\cal A}_2$, both for educational purposes and for practical ones, since one can think of an experiment to measure the ${\cal A}_2$ distribution, i.e. by means of an additional optical field \cite{Gershnabel1}.
Fig. \ref{StaticSmallField_NoPre} is very similar to Fig. $5$ from \cite{Gershnabel1} (in both cases the $D$'s values are of the same order of magnitude). Therefore, the interpretation of the results given in \cite{Gershnabel1} is applicable to the present case too. Briefly, the tail of the distribution is formed by relatively low-energy molecules that are angularly trapped by the deflecting field  (${\cal A}_1$ and ${\cal A}_2$ are close to $1$). Some of the molecules in the ensemble are rotating in vertical planes that contain the deflecting field polarization. They  contribute to the negative peak of Fig. \ref{StaticSmallField_NoPre}a, and the rainbow peak in Fig. \ref{StaticSmallField_NoPre}b around the value ${\cal A}_2=0.28$ (for slightly trapped molecules) and ${\cal A}_2=0.5$ (for free molecules). The main difference between Fig. \ref{StaticSmallField_NoPre}b and Fig. $5$b from \cite{Gershnabel1} is the suppressed peak value at ${\cal A}_2=0.5$ in Fig. \ref{StaticSmallField_NoPre}b, as well as the less clear peak at ${\cal A}_2=0.28$. The suppression of these peaks is due to the important difference between linear and symmetric molecules, namely the possibility of the symmetric molecules to rotate around their own symmetry axis. Therefore, unlike the case of linear molecules, which rotate both with and against the electric field, a portion of the symmetric molecules in the ensemble may interact only along the symmetric field (and hence be more deflected by it).
We also examine the case of a stronger deflection, i.e. by means of a $2\times 10^8 V/m$ deflecting field, where $C=-9.7\times 10^{-3}$ and $D=8$. The results are presented in Fig. \ref{StaticHighField_NoPre}, where a larger portion of the molecules is trapped by the electric field. This Fig. may be compared to the quite similar Fig. $6$ from \cite{Gershnabel1}.
Finally, we apply a prealigning laser pulse ($P=25$) polarized parallel to the direction of the deflecting field (i.e. in the $Z$ direction). The results are given in Fig. \ref{StaticHighField_Pre25} and are very similar to the results presented in Fig. $8$ of \cite{Gershnabel1}. Due to the initial prealignment, the molecules gain a strong rotational energy and the interaction with the deflecting field is almost cancelled out (Fig. \ref{StaticHighField_Pre25}a). Furthermore, the rainbow at ${\cal A}_2=0.5$ is enhanced due to the strongly rotating molecules in the vertical planes.
Thus, we show that by means of laser-induced prealignment, a dramatic control of the scattering distribution is possible. In particular, the interaction between the molecule and the static field can be switched-off (and on).

\section{Conclusions} \label{Conclusions}

We considered scattering of symmetric molecules by external optical and static electric fields, and found new singularities in the distribution of the scattering angle. Depending on the shape of the molecule, i.e. oblate/prolate geometry, the singularities were shown to be suppressed or enhanced compared to the case of linear molecules. We developed a simple model for the scattering process, and found a good correspondence between its predictions and the ones of more complicated treatments, including adiabatic invariants formalism and direct simulation of the  Euler-Lagrange equations of motion. Our results indicate that laser-induced prealignment provides an effective tool for controlling the deflection of the molecules, and it may be used for increasing the brightness of the scattered molecular beam. Furthermore the interaction between the molecules and a static electric field can be switched-off (or switched-on) by means of the prealignment. This may be important for nanofabrication schemes based on the molecular optics approach \cite{Seideman}. Moreover, molecular deflection by nonresonant optical dipole force is considered a promising route to separation of molecular mixtures (for a recent review, see Ref. \cite{Zhao}). Narrowing the distribution of the scattering angles may substantially increase the efficiency of separation of multicomponents beams, especially when the prealignment is applied selectively to certain molecular species, such as isotopes \cite{Fleischer} or nuclear spin isomers \cite{water Gershnabel,Renard,Fleischer1}. The predictions of our theory may encourage others to design more complicated schemes for controlling the process, e.g. by using several prealignment pulses with variable polarizations. Future studies may also focus on the scattering of asymmetric molecules having  more complicated deflection potential. Moreover, the same mechanisms may prove efficient for controlling inelastic molecular scattering off metalic and dielectric surfaces. These and other aspects of the present problem are subjects of an ongoing investigation.

\section*{ACKNOWLEDGMENTS}

We acknowledge the support of our study by a grant from
the Israel Science Foundation. This research is made possible
in part by the historic generosity of the Harold Perlman family.
I.A. is an incumbent of the Patricia Elman Bildner Professorial
Chair.

\bibliographystyle{phaip}

\begin{references}

\bibitem{Wayne} R. Velotta et al., Phys. Rev. Lett. \textbf{87}, 183901 (2001); J. Itatani
et al., ibid. \textbf{94}, 123902 (2005).

\bibitem{Bartels} R. A. Bartels et al., Phys. Rev. Lett. \textbf{88}, 013903 (2002); V.
Kalosha et al., ibid. \textbf{88}, 103901 (2002).

\bibitem{Gordon} R. J. Gordon, L. Zhu, W. A. Schroeder, and T. Seideman, J.
Appl. Phys. \textbf{94}, 669 (2003).

\bibitem{Larsen} J. J. Larsen, I. Wendt-Larsen, and H. Stapelfeldt, Phys. Rev.
Lett. \textbf{83}, 1123 (1999); M. Tsubouchi et al., ibid. \textbf{86}, 4500
(2001); I. V. Litvinyuk et al., ibid. \textbf{90}, 233003 (2003).

\bibitem{Shapiro} E. A. Shapiro, M. Spanner, and M. Y. Ivanov, Phys. Rev. Lett.
\textbf{91}, 237901 (2003); J. Mod. Opt. \textbf{52}, 897 (2005); K. F. Lee et
al., Phys. Rev. Lett. \textbf{93}, 233601 (2004).

\bibitem{Heritage} J. P. Heritage, T. K. Gustafson, and C. H. Lin, Phys. Rev. Lett.
\textbf{34}, 1299 (1975); J. Ortigoso et al., J. Chem. Phys. \textbf{110}, 3870
(1999); T. Seideman, Phys. Rev. Lett. \textbf{83}, 4971 (1999); L. Cai,
J. Marango, and B. Friedrich, ibid. \textbf{86}, 775 (2001).

\bibitem{Stapelfeldt_symm} H. Stapelfeldt, Eur. Phys. J. D \textbf{26}, 15 (2003).

\bibitem{Hamilton_symm} E. Hamilton et al, Phys. Rev. A \textbf{72}, 043402 (2005).

\bibitem{Kitano} K. Kitano, H. Hasegawa, and Y. Ohshima, Phys. Rev. Lett. \textbf{103}, 223002 (2009).

\bibitem{Pabst} S. Pabst, and R. Santra, Phys. Rev. A \textbf{81}, 065401 (2010).

\bibitem{Gershnabel} E. Gershnabel, and I. Sh. Averbukh, Phys. Rev. Lett. \textbf{104}, 153001 (2010); Phys. Rev. A \textbf{82}, 033401 (2010).

\bibitem{Gershnabel1} E. Gershnabel, and I. Sh. Averbukh, J. Chem. Phys. \textbf{134}, 054304 (2011).

\bibitem{Floss} J. Flo\ss, E. Gershnabel, and I. Sh. Averbukh, Phys. Rev. A \textbf{83}, 025401 (2011).

\bibitem{Stapelfeldt} H. Stapelfeldt et al., Phys. Rev. Lett. \textbf{79}, 2787 (1997); H.
Sakai et al., Phys. Rev. A \textbf{57}, 2794 (1998).

\bibitem{Bum} Bum Suk Zhao et al., Phys. Rev. Lett. \textbf{85}, 2705 (2000);
Hoi Sung Chung et al., J. Chem. Phys. \textbf{114}, 8293 (2001);
Bum Suk Zhao et al., J. Chem. Phys. \textbf{119}, 8905 (2003).

\bibitem{Purcell} S. M. Purcell, and P. F. Barker, Phys. Rev. Lett. \textbf{103}, 153001
(2009); Phys. Rev. A {\bf 82}, 033433 (2010).

\bibitem{Landau} L. D. Landau and E. M. Lifshitz, Mechanics, 3rd ed.
(Butterworth-Heinemann, Oxford, 1976).

\bibitem{Seideman} T. Seideman, J. Chem. Phys. \textbf{106}, 2881 (1997); \textbf{107}, 10420 (1997); \textbf{111}, 4397 (1997).

\bibitem{water Gershnabel} E. Gershnabel, and I. Sh. Averbukh, Phys. Rev. A \textbf{78}, 063416 (2008).

\bibitem{Dugourd} P. Dugourd, I. Compagnon, F. Lepine, R. Antoine, D. Rayane,
and M. Broyer, Chem. Phys. Lett. \textbf{336}, 511 (2001).

\bibitem{Goldstein} H. Goldstein, C. Poole, and J. Safko, Classical Mechanics, 3rd
ed. (Addison-Wesley, Reading, MA, 2001).

\bibitem{Hasegawa} H. Hasegawa, and Y. Ohshima, Chem. Phys. Lett. \textbf{454}, 148 (2008).

\bibitem{Hartinger} K. Hartinger, and R. A. Bartels, J. Opt. Soc. Am. B \textbf{25}, 407 (2008).

\bibitem{Ito} F. Ito, J. Chem. Phys. \textbf{124}, 054309 (2006).

\bibitem{Meerts} W. L. Meerts and I. Ozier, J. Chem. Phys. \textbf{75}, 596 (1981).

\bibitem{Zhao} B. S. Zhao, Y. M. Koo, and D. S. Chung, Analytica Chimica Acta \textbf{556}, 97 (2006).

\bibitem{Fleischer} S. Fleischer, I. Sh. Averbukh, and Y. Prior, Phys. Rev. A \textbf{74}, 041403(R) (2006).

\bibitem{Renard} M. Renard, E. Hertz, B. Lavorel, and O. Faucher, Phys. Rev. A \textbf{69}, 043401 (2004).

\bibitem{Fleischer1} S. Fleischer, I. Sh. Averbukh, and Y. Prior, Phys. Rev. Lett. \textbf{99}, 093002 (2007).

\end{references}

\end{document}